\documentstyle[11pt]{article}
\input amssymb.sty

\title{A Defense of the Paraconsistent\\Approach to Quantum Superpositions\\(Answer to Arenhart and Krause)}

\author{{\sc Christian de Ronde}\thanks{Fellow Researcher of the Consejo
Nacional de Investigaciones Cient\'{\i}ficas y T\'ecnicas. Email: cderonde@vub.ac.be}}
\date{}

\begin{document}

\bibliographystyle{plain}
\maketitle

\begin{center}
\begin{small}
Philosophy Institute ``Dr. A. Korn" \\ 
Buenos Aires University, CONICET - Argentina \\
Center Leo Apostel and Foundations of  the Exact Sciences\\
Brussels Free University - Belgium \\
\end{small}
\end{center}

\begin{abstract}
\noindent In \cite{daCostadeRonde13}, Newton da Costa together with the author of this paper argued in favor of the possibility of considering quantum superpositions in terms of a paraconsistent approach. There we argued that, even though most interpretations of Quantum Mechanics (QM) attempt to escape contradictions, there are many reasons that indicate it could be worthwhile to engage in a research of this kind. Recently, Arenhart and Krause \cite{ArenhartKrause14} have raised several arguments against this approach. In the present paper we attempt to answer the main questions and obstacles presented by them. We will argue, firstly, that the obstacles presented by Arenhart and Krause are based on a specific metaphysical stance, which we will characterize in terms of what we call the Orthodox Line of Research (OLR). Secondly, that this is not necessarily the only possible line, and that a different one, namely, a Constructive Metaphysical Line of Research (CMLR) provides a different perspective in which  the Paraconsistent Approach to Quantum Superpositions (PAQS) can be regarded as a valuable prospect that could be used by different interpretations of quantum mechanics. Finally, we provide a set of specific answers to the main problems raised by Arenhart and Krause in order to clarify our line of research as well as the original perspective introduced by the PAQS.
\end{abstract}
\begin{small}


\end{small}

\bibliography{pom}

\newtheorem{theo}{Theorem}[section]

\newtheorem{definition}[theo]{Definition}

\newtheorem{lem}[theo]{Lemma}

\newtheorem{met}[theo]{Method}

\newtheorem{prop}[theo]{Proposition}

\newtheorem{coro}[theo]{Corollary}

\newtheorem{exam}[theo]{Example}

\newtheorem{rema}[theo]{Remark}{\hspace*{4mm}}

\newtheorem{example}[theo]{Example}

\newcommand{\proof}{\noindent {\em Proof:\/}{\hspace*{4mm}}}

\newcommand{\qed}{\hfill$\Box$}

\newcommand{\ninv}{\mathord{\sim}} %involutive negation

\newtheorem{postulate}[theo]{Postulate}

\section*{Introduction}

In \cite{daCostadeRonde13}, Newton da Costa together with the author of this paper argued in favor of the possibility of considering quantum superpositions in terms of a paraconsistent approach. We argued that, even though most interpretations of Quantum Mechanics (QM) attempt to escape contradictions, there are many reasons why it could be worthwhile to engage in a research of this kind. Recently, Arenhart and Krause \cite{ArenhartKrause14} have raised several arguments against the Paraconsistent Approach to Quantum Superpositions (PAQS). Their obstacles are condensed in the following six main statements: 

\begin{enumerate}
\item[I.] The PAQS does not allow for contradictions due to its incompatibility with the Semantic Requirement, the Minimal Property Ascription Condition and the Paraconsistent Property Ascription. 
\item[II.] The PAQS obscures the meaning of probability in QM.
\item[III.] The PAQS does not explain the measurement problem.
\item[IV.] Contrary to what is claimed by PAQS, contradictions are not observed in QM.
\item[V.]  The PAQS inflates unnecessarily the population of the world with contradictions.
\item[VI.] The PAQS does not explain the vanishing of terms in the superposition after measurement. 
\end{enumerate}

First of all we must remark that in \cite{daCostadeRonde13} we did not propose an interpretation nor a metaphysical scheme but only called the attention to the need of considering quantum superpositions as ontologically robust and the possibility of doing so in terms of a paraconsistent approach. However, including contradictions might be regarded as preparing the stage for a metaphysical step, this is why we believe that ---now entering the scenery of metaphysics--- the remarks and considerations of Arenhart and Krause deserve careful attention as well as answers. In this paper we will argue that:\\

\noindent {\it i) Arenhart and Krause place their obstacles from a specific metaphysical stance, which we will characterize in terms of what we call the Orthodox Line of Research (OLR).\\

\noindent ii) That this is not necessarily the only possible line, and that a different one, namely, a Constructive Metaphysical Line of Research (CMLR) provides a different perspective in which PAQS can be regarded as a valuable prospect. Furthermore, that within the CMLR the problems and obstacles raised by Arenhart and Krause disappear.}\\ 

\noindent Firstly, we will characterize the OLR in order to show later on how the arguments raised by Arenhart and Krause rely on this particular metaphysical stance. We will also put forward an alternative line, the CMLR, which implies a different metaphysical stance in order to confront the problem of interpreting QM in general, and quantum superpositions in particular. We will argue that the OLR and the CMLR determine a different set of problems and interpretational strategies. In the second section of this paper we will analyze the need, implied by the CMLR, of bringing onto stage a different metaphysical scheme to the one assumed by the OLR. We will discuss the famous `measurement problem' of the OLR and argue that, from the perspective of the CMLR this problem can be inverted and turned into what we call `the superposition problem'. In section 3, we discuss what we know about quantum superpositions and analyze the meaning of contradiction in QM as related to the formal understanding of quantum possibility and its contextual constraints as exposed in the Modal Kochen-Specker theorem. In section 4, we give specific answers to the six main obstacles mentioned above. Finally, in the last section, we present some remarks and analyze the proposal of Krause and Arenhart to understand contradictions in terms of the square of opposition.

\section{Metaphysical Stances, Interpretational Strategies and Problems}

In {\it The Empirical Stance}, Bas van Fraassen \cite[p. xviii]{VF02} makes a remarkable claim: ``The problem of appearance and reality affects first of all philosophy itself. I argue for a view of philosophy as a stance, as existential.'' To consider philosophy as a stance, as existential, means that our analysis is not void of values, intentions and presuppositions. These must be made explicit in order to be honest about the limits of our own arguments.  From this perspective ---which we support--- quite immediately we are driven into the specific consideration of the metaphysical presuppositions that we accept or not in a given analysis. Physics does not escape this state of affairs, as remarked by Einstein \cite[p. 1196]{PS}:  ``The problem is that physics is a kind of metaphysics; physics describes `reality'. But we do not know what `reality' is. We know it only through physical description...'' Indeed, due to the fact a physical problem is always constituted through a metaphysical perspective there is no neutral nor distant questioning regarding a physical object. Posing a problem involves definite metaphysical presuppositions and choices without which questions remain meaningless. The problem constitutes the object of inquiry itself. The problem of interpreting QM, for example, which makes complete sense from a realist perspective is completely meaningless from an instrumentalist one. Instrumentalism has a set of problems different from those of realism. Furthermore, a problem determines ---implicitly--- its own set of possible answers, placing in itself a limit to knowledge and understanding and it is through this same limit that we learn about the world. As we shall see, posing problems to QM is a subtle task which is some times betrayed by hidden agendas.

\subsection{The OLR (and its Problems)}

Let us first start by characterizing the OLR. We can do so in terms of two main metaphysical presuppositions which seem to have sedimented in present philosophy of QM. Both presuppositions can be found to play a major role not only in most interpretations of QM but also in the problems posed in the literature. The first one is related to a widespread idea concerning the need to unify different physical theories. As we have discussed in \cite{deRonde11}, this is not necessarily the only possible way to understand the relation between theories. Nevertheless this metaphysical ideal has imposed the need to find a bridge between the quantum formalism and the classical physical representation of the world. This first presupposition was already stated by Bohr in terms of his {\it correspondence principle} \cite{BokulichCP}.

\begin{enumerate}
\item {\bf Quantum to Classical Limit:}  The principle that one can find a  bridge between classical mechanics and QM; i.e., that the main notions of classical physics can be used in order to explain quantum theory.
\end{enumerate}

The second metaphysical principle which has guided the OLR can be also traced to Bohr's claim that physical experience needs to be expressed exclusively in terms of classical language \cite{BokulichBokulich}. According to Bohr \cite[p. 7]{WZ}: ``[...] the unambiguous interpretation  of any measurement must be essentially framed in terms of classical physical theories, and we may say that in this sense the language of Newton and Maxwell will remain the language of physicists for all time.'' Furthermore \cite[p. 7]{WZ} ``it would be a misconception to believe that the difficulties of the atomic theory may be evaded by eventually replacing the concepts of classical physics by new conceptual forms.''

\begin{enumerate}
\item[2.] {\bf Classical Physical Representation:} The principle that one needs to presuppose in any interpretation of QM the representation provided by classical physics in terms of the entities with definite actual properties.
\end{enumerate}

\noindent If one considers the core of the classical (Newtonian) physical and metaphysical representation of the world\footnote{See for discussion \cite{RFD14}.} one is then stuck with two main concepts: `entity' and `actuality' (as a mode of existence). In QM one can also encounter these metaphysical notions as basic elements of any interpretation.
Let us make a short detour that will help us to understand not only the importance of these concepts but also their historicity and development ---to which we shall return through the rest of the paper. The notions of `entity' and `actuality' were created by Aristotle. Also `potentiality' played a major role in Aristotle's metaphysics: movement was described in terms of the path from potentiality, which contained the undetermined, contradictory and non-individual realm of existence, to the actual mode of being, determined through the logical principle of existence (PE), the principle of non-contradiction (PNC) and the principle of  identity (PI). These same principles allowed Aristotle to put forward ---together with classical logic itself (see for discussion \cite{VerelstCoecke})--- the notion of entity. The notion of entity was capable of unifying, of totalizing in terms of a ``sameness'', creating certain stability for knowledge to be possible. This representation or transcendent description of the world is considered by many as the origin of metaphysical thought itself. Actuality is then linked directly to metaphysical representation and understood as characterizing a mode of existence independent of observation. Indeed, this is the way through which metaphysical thought was able to go beyond the  {\it hic et nunc}, creating a world beyond the world, a world of concepts. In physics, it was this possibility which allowed us to deal with experience in a counterfactual manner and predict even non-performed experiments. 

The transition from medieval to modern science coincides with the abolition of Aristotelian hylomorphic metaphysical scheme ---in terms of potentiality and actuality--- as the foundation of knowledge. However, the basic structure of Aristotelian logic still remained the basis for correct reasoning \cite[p. 7]{VerelstCoecke}. As a consequence, potentiality was completely eliminated from physics. After Newton, only actuality was considered in order to account for the mode of existence of physical objects.  Indeed, in classical mechanics the representation of the state of the physical system is given by a point in phase space $\Gamma$ and the physical magnitudes are represented by real functions over $\Gamma$. These functions commute and can be interpreted as possessing definite (non-contradictory) values independently of measurement, i.e. each function can be interpreted as being actual. The term actual refers in this case to {\it preexistence} (within the transcendent representation) and not to the observation {\it hic et nunc}. Every physical system may be described exclusively by means of its actual properties and its evolution, through the evolution of such (actual) properties. Thus, potential or possible properties are considered only as the points to which the system might arrive in a future instant of time. The physics of Newton became a physics of pure actuality, providing a description of the universe in terms of an {\it Actual State of Affairs} (ASA) \cite[p. 124]{Dieks10}. 

Going back to QM, the OLR is based implicitly on this metaphysical scheme and seeks for answers in terms of the set of (metaphysical) presuppositions derived from it. But from a methodological perspective it seems not very smart to take as standpoint, in order to solve problems, presuppositions which ---we already know--- have problems with the theory in the first place. If everything seems to point in the direction that QM has problems with the notions of `individuality', `definiteness', `space', `time', `entity', `actuality', etc. Why should we keep trying to solve problems with these very same concepts? Why should we accept that the only way of posing problems is according to the classical metaphysical scheme? As remarked already by Dieks \cite[p. 1417]{Dieks89a}: ``This would deny the possibility of really new fundamental theories, conceptually independent of classical physics.'' Due to the closure of the scheme, the problems posed by the OLR concentrate on {\it justifying} the classical description of the world in terms of entities constituted by properties in the actual mode of existence.\footnote{We can find many examples of such problems in the literature: the quantum to classical limit, which was supposedly resolved through decoherence, the problem of non-locality which implicitly considers space-time in relation to QM, the problem of identical particles which presupposes the notion of entity and the problem of holism and quantum separability which also assumes, as a standpoint, that we have quantum systems and that we can `cut' such systems into parts.} One of the main problems that interests us here and will be subject of analysis and discussion is the famous measurement problem (Section 2).

\subsection{The CMLR (and its Problems)}

We have analyzed in \cite{deRonde10} how to separate between two main strategies regarding the problem of interpreting QM. The first strategy consists in beginning with a presupposed set of classical metaphysical principles and advance towards a new formalism that is able to account for such principles. Examples of this strategy are the collapse theory proposed by Ghirardi, Rimini and Weber \cite{GRW} which introduces non-linear terms in the Schr\"odinger equation and Bohmian mechanics which introduces space-time particles \cite{Bohm53}.\footnote{As remarked by Bitbol \cite[p. 8]{Bitbol10}: ``Bohm's original theory of 1952 is likely to be the most metaphysical (in the strongest, speculative, sense) of all readings of quantum mechanics. It posits free particle trajectories in space-time, that are unobservable in virtue of the theory itself."} The second strategy consists in accepting the orthodox formalism of QM and advance towards the creation and elucidation of the metaphysical principles which would allow us to answer the question: what is QM talking about? Examples of this second strategy are quantum logic and its different lines of development such as the Geneva School of Jauch and Piron \cite{Piron76} or the modal interpretation of Van Fraassen and Dieks \cite{Vermaas99}. From this perspective, the importance is to focus on the formalism of the theory and try to learn about the symmetries, the logical features and structural relations. The idea is that, by learning about such aspects of the theory we can also develop the metaphysical conditions that should be taken into account in a coherent interpretation of QM. However, even within this second strategy which seems less keen to embrace classical metaphysics, the OLR is very strong. As a matter of fact, many approaches which take the formalism as a standpoint end up going back to the classical metaphysical notions that we have already characterized. Our proposal is to consider the second strategy but accept the possibility to go beyond the classical metaphysical representation ---in terms of entities or an ASA---, engaging at the same time in the necessary  creation of new concepts in order to find a suitable (non-classical) metaphysical scheme that would allow us to interpret coherently QM and its phenomena. 

Although we understand that the OLR was, at the beginning, the most reasonable and obvious path to follow ---mainly due to its success over the three centuries before the creation of QM---, time has proven that this line of research has not been a very successful one for interpreting QM. As a matter of fact, after more than one century, we have not seemed to advance very much in the understanding of the theory nor have we been able to develop a coherent interpretation. Still today, it is safe to say we do not know what QM is talking about.

As we remarked already, regarding the quantum to classical limit principle of the OLR there exist many philosophical positions which do not assume the need of a {\it limit} between physical theories. Following Heisenberg we have argued elsewhere in favor of a closed theory approach \cite{deRonde11, deRonde14}. As remarked by Bokulich \cite[p. 79]{Bokulich04}: ``The German phrase that Heisenberg uses is {\it abgeschlossene Theorie}, where {\it abgeschlossene} can be translated as `closed', `locked', `isolated', or `self-contained'.'' Heisenberg understands `closed theories' as a relation of tight interconnected concepts, definitions and laws whereby a large field of phenomena can be described.  Every physical theory needs to develop its own conceptual scheme.\footnote{As remarked by Heisenberg in an interview by Thomas Kuhn \cite[p. 98]{Bokulich06}: ``The decisive step is always a rather discontinuous step. You can never hope to go by small steps nearer and nearer to the real theory; at one point you are bound to jump, you must really leave the old concepts and try something new... in any case you can't keep the old concepts.''} The radical incommensurability assumed by the closed theory approach puts an important restriction to the metaphysical assumption of a necessary limit between different theories. The only important aspect to consider a physical theory as `closed' is the internal {\it coherency} between the formal mathematical elements, the conceptual structure and the physical experience created by these same concepts. QM need not be talking about the same as classical mechanics. An important aspect of this approach is that each physical theory is only able to attack a restricted set of problems and questions, those which presuppose the concepts and formal structure put forward by the theory itself.

Taking into account the need to provide a coherent physical interpretation of QM, our CMLR is based on three main presuppositions already put forward and discussed in \cite[pp. 56-57]{deRonde11}.

\begin{enumerate}
\item {\bf Closed Representational Stance:} Each physical theory is closed under its own formal and conceptual structure providing access to a specific set of phenomena. The theory provides the constraints to consider, explain and understand physical phenomena. 

\item {\bf Formalism and Empirical Adequacy:} The formalism of QM is able to provide (outstanding) empirically adequate results. Empirical adequacy determines the success of a theory and not its (metaphysical) commitment to a certain presupposed conception of the world. The problem is not to find a new mathematical scheme, on the contrary, the `road signs' point in the direction that {\it we must stay close to the orthodox quantum formalism}.

\item {\bf Constructive Stance:} To learn about what the formalism of QM is telling us about reality we might be in need of {\it creating new physical concepts}.
\end{enumerate}

\noindent What is needed according to the CMLR is a radical inversion of orthodoxy and its problems. According to this inversion the features of QM should be all considered instead as problems, as the main characteristics that must be considered in the development of a coherent interpretation of the theory.

\section{The Superposition Problem: Beyond Entities and Actuality}
 
The measurement problem is one of the main questions imposed by the OLR. This problem can be clearly stated in the following manner:\\ 

\noindent {\it {\bf Measurement Problem (MP):} Given a specific basis (or CSCO) QM describes mathematically a state in terms of a superposition (of states), since the evolution described by QM allow us to predict that the quantum system will get entangled with the apparatus and thus its pointer positions will also become a superposition, the question is why do we observe a single outcome instead of a superposition of them?}\\
 
\noindent As we have extensively discussed in \cite{deRonde15b} this problem shouldn't be confused with the basis problem nor answered by arguing that: in a different basis in which the $\Psi$ is written as a superposition of one single term the outcome is not a superposition of pointers but one single pointer ``right from the start''. This is changing the subject of inquiry through a change of basis which is also problematic in QM due to the contextual character of the theory. If we ask about the interpretation of a quantum superposition: $c_{1x} | \uparrow_x \rangle + \ c_{2x} | \downarrow_x \rangle$, given the situation in which a Stern Gerlach apparatus is in the $x$-direction, the answer cannot be that if we change the direction of the Stern Gerlach to the $j$-direction the superposition will be a single term $| \uparrow_j \rangle$ and thus interpreted in terms of actual state of affairs. 

Although the MP accepts the fact that there is something very weird about quantum superpositions, leaving aside their problematic meaning, it focuses on the justification of the actualization process. Taking as a standpoint the single actual result it asks: how do we get there from the quantum superposition? The questioning is completely analogous to the one posed by the quantum to classical limit problem: how do we get from weird QM into our common sense classical physical description of the world? The MP is thus an attempt to justify why, regardless of QM, we only observe actuality. The problem places the result in the origin, what needs to be justified is the already known answer. But, could it be we were asking the wrong question?

According to the CMLR we need to see the problem from a different perspective, we need to think differently. Our approach has attempted to escape the ruling of the notion of actuality  ---put forward by the actualist metaphysical scheme of Newtonian mechanics (Section 1.1)--- by inverting the MP, turning upside-down the focus, concentrating on the physical meaning of quantum superpositions instead of trying to justify what we already know \cite{deRonde11}. Most interpretations, focusing completely on the MP, have remained completely silent regarding the physical representation of superpositions. We believe that the answer to this problem might be the key to truly understand the latest technical and experimental developments done today in laboratories around the world \cite{Nature13, Nature11a, NaturePhy12, Nature07}. But before stating the problem some remarks go in order. Firstly, given a $\Psi$ we call a quantum superposition to any mathematical representation provided in terms of a specific basis (or context). Second, the MP is context dependent, it needs a specific basis to be analyzed, if we change the basis to one in which the $\Psi$ is written as one single term the problem is not being discussed. Third, what is interesting is how to interpret the most general case of quantum superpositions, namely, those superpositions in which we have more than one term. In particular, the superpositions that have a deep difficulty in being interpreted are those of the type of a Schr\"odinger cat, e.g. $c_{1x} | \uparrow_x \rangle + \ c_{2x} | \downarrow_x \rangle$, which are composed by a property and its contradictory (see for discussion \cite{daCostadeRonde13}). Having said this we can now state the Superposition Problem in the following terms:\\ 

\noindent {\it {\bf Superposition Problem (SP):} Given a situation in which there is a quantum superposition of more than one term, $\sum c_i \ | \alpha_i >$, and given the fact that each one of the terms relates trough the Born rule to a meaningful physical statement,\footnote{In \cite{deRonde15b, deRonde15c} we have defined this notion in the following terms. {\bf Meaningful Physical Statements (MPS):} {\it If given a specific situation a theory is capable of predicting in terms of definite physical statements the outcomes of possible measurements, then such physical statements are meaningful to the theory and must be constitutive parts of the particular representation of physical reality that the theory provides. Measurement outcomes must be considered only as an exposure of the empirical adequacy (or not) of the theory.}} the problem is how do we physically represent this mathematical expression, and in particular, the multiple terms?}\\
 
It should be noticed that the outstanding technological and experimental developments of the last decades is based in superpositions of more than term and not in the particular superposition of one single term which is the only one orthodoxy is able to interpret. While the MP focuses in the explanation of the measurement outcome, the SP concentrates in providing a physical representation of quantum superpositions themselves. Indeed, as Heisenberg makes the point \cite[p. 264]{Heis73}: ``The history of physics is not only a sequence of experimental discoveries and observations, followed by their mathematical description; it is also a history of concepts. For an understanding of the phenomena the first condition is the introduction of adequate concepts. Only with the help of correct concepts can we really know what has been observed.'' Before we can understand the process of actualization by interpreting the Projection Postulate (PP) we first need to find a physical concept which allows us to picture what a quantum superposition {\it is} or {\it represents}; for there is an obvious asymmetry in comparing, on the one hand, a mathematical expression and, on the other, an actualized outcome.\footnote{It should be clear that  there is no self evident path between the superposition and its outcome. As it is well known there are multiple ways of interpreting the PP. See for a discussion \cite{RFD14}.}  

Let us, for the sake of the argument, grant that the CMLR is the correct path to follow ---which remains of course a logical possibility. How should we then proceed? The perspective is quite different from the OLR regarding what needs to be done in order to interpret QM. If we consider it as a closed theory we should not concentrate on its relation to classical physics but rather attempt to develop QM itself ---and only later on try to find out how QM relates to classical physics. In such case we would need to find new concepts which match the formalism in analogous fashion to the way the notions of object, space and time allow us to interpret the evolution of a mathematical point within the mathematical formalism of classical mechanics. 

The logical and ontological PE, PNC and PI created by Aristotle allowed Newton to develop a metaphysical representation of the physical world based on the notions of entity and actuality (Section 1.1). QM seems to have many difficulties to be interpreted following this formal and conceptual scheme. As a matter of fact QM has problems with all three Aristotelian metaphysical principles. It has problems with the PE due to the contextual character of the theory which precludes the global valuation of the properties of a quantum system \cite{KS, Peres}, it has problems with the PI due to the question of quantum individuality \cite{FrenchKrause} and it also seems to have problems with the PNC due to the existence of superpositions \cite{daCostadeRonde13}. But just in the same way as non reflexive logics might help us to understand the meaning of quantum individuals, or that dynamical logics \cite{BaltagSmets12} and category theory \cite{CoeckeHeunenKissinger13} might allow us to better understand quantum interactions, the PAQS opens the door to the possibility of considering physical notions which are not restricted to the logical PNC. The history of physics is full of developments directly linked to formal shifts. For example, non-euclidean geometry was a key formal development which allowed not only to produce new notions of  space and time, but also to create the theory of relativity itself. In the same way, the introduction of  paraconsistent logics might provide a valuable help to investigate the physical meaning of quantum superpositions. There are some new formal proposals which following this path are starting to seriously explore the possibilities of such logical approach \cite{daCostadeRonde14, KrauseArenhart14}.

\section{Superpositions, Potentiality and Contradictions in Quantum Mechanics}

Fortunately, experimentalists (in actual laboratories!) do not seem to care much about philosophical discussions regarding QM. Quite independently of the MP they have kept using quantum superpositions and the orthodox formalism in order to  produce the most outstanding technical developments of the last decades. But although we can use superpositions to teleport information or implement quantum computers, we still cannot find the physical concept which unifies all we have learnt about them. Indeed, there are many characteristics and behaviors we have learnt about superpositions: we know about {\it their existence regardless of the effectuation of one of its terms}, as shown, for example, by the interference of different possibilities in {\it welcher-weg} type experiments \cite{ Nature11a, NaturePhy12}, {\it their reference to contradictory properties}, as in Schr\"{o}dinger cat states \cite{Nature07}, we also know about {\it their non-standard route to actuality}, as explicitly shown by the MKS theorem \cite{RFD14, DFR06}, and we even know about {\it their non-classical interference with themselves and with other superpositions}, used today within the latest technical developments in quantum information processing \cite{Nature13}. In spite of the fact we still cannot say what a quantum superposition {\it is} or {\it represents}, we must admit that they seem ontologically robust.  

It should be clear that the importance to find out how to physically represent quantum superpositions is not only philosophical but also technological for it is only through physical concepts that really new experiments can be designed. There are many elements which can be seen as ``road signs'' that point in the direction of an ontological interpretation of quantum superpositions. If the terms within a quantum superposition are considered as quantum possibilities (of being actualized) ---and it seems difficult not to agree with such an idea--- then we must also admit that such {\it possibilities interact} according to the Schr\"odingier equation. It is also well known that one can produce {\it interactions between multiple superpositions} (entanglement) and then calculate the {\it evolution of all terms} as well as predict the ratio of all probable outcomes. It then becomes difficult not to believe that these terms that `interact', `evolve' and `can be predicted' according to the theory, are not (in some way) real. This is the main reason, which we find very strong, to interpret all terms in the superposition as existing (in some way). Since we know that each term in the superposition relates to a specific possibility ---which can interact with a different possibility--- it makes sense to develop an ontological notion of possibility which supports whatever quantum superpositions are. But, exactly because of what we have learnt already, we should be careful not to claim ---as Arenhart and Krause seem to imply--- that superpositions exist in the realm of actuality. 
 
We believe a reasonable strategy would then be to start with what we know works perfectly well, namely, the orthodox formalism of QM.  Starting from the formalism, a good candidate to develop a mode of existence is of course {\it quantum possibility}. In several papers, together with Domenech and Freytes, we have analyzed how to understand possibility in the context of the orthodox formalism of QM \cite{DFR06, DFR08a, DFR08b, DFR09}. From this investigation there are several conclusions which can be drawn. We started our analysis with a question regarding the contextual aspect of possibility. As it is well known, Kochen-Specker (KS) theorem does not talk about probabilities, but rather about the constraints of the formalism to actual definite valued properties considered from multiple contexts \cite{KS}. What we found via the analysis of possible families of valuations is that a theorem which we called ---for obvious reasons--- the Modal KS (MKS) theorem can be derived which proves that quantum possibility, contrary to classical possibility, is also contextually constrained. This means that, regardless of its use in the literature: quantum possibility is not classical possibility.\footnote{For a discussion regarding the important distinction between mathematical formalism and physical interpretation, as well as between the algebraic structure, the language and the meta-language, see: \cite[p. 15]{RFD14}.} In a recent paper, \cite{RFD14} we have concentrated on the analysis of the PP within the orthodox frame and interpreted, following the structure, the logical realm of possibility in terms of potentiality.

Once we accept we have two distinct realms of existence, namely, potentiality and actuality, we must be careful about the way in which we define {\it contradictions}. Certainly, contradictions cannot be defined in terms of truth valuations in the actual realm, simply because we have distinguished that the notion that must interpret superpositions is an existent in the potential realm ---not in the actual one. The MKS theorem shows explicitly that a quantum wave function implies multiple incompatible valuations which can be interpreted as {\it potential contradictions} \cite{RFD14}. Thus, one can claim that while contradictions exist in the potential realm, they can never be found in actuality. Our analysis has always advocated the idea that contradictions ---by definition--- are never found in the actual realm. Our attempt is to turn things upside-down: we do not need to explain the actual via the potential but rather, we need to use the actual in order to develop the potential \cite[p. 148]{deRonde11}.

\section{Answers to Arenhart and Krause}

Now that we have specified our perspective, as well as the stance implicitly assumed by Arenhart and Krause, we are ready to address more specifically some of the arguments and obstacles raised by them against the PAQS.\\

\noindent{\bf I. The PAQS does not allow for contradictions due to its incompatibility with the Semantic Requirement (SR), the Minimal Property Ascription Condition (MPAC) and the Paraconsistent Property Ascription (PPA).}

\noindent In order to criticize the notion of contradiction supposedly assumed by the PAQS, Arenhart and Krause \cite{ArenhartKrause14} present several conditions which, they claim, become incompatible within the approach. Firstly they propose the SR: ``Contradictory statements [of the language] must have opposite truth values.'' To analyze a specific situation in QM they take the following quantum state: $\alpha  \ | \uparrow_x > +  \  \beta \ | \downarrow_x >$ and claim that: ``The statements corresponding to properties represented by $| \uparrow_x > < \uparrow_x |$ and $| \downarrow_x > < \downarrow_x |$ must have opposite truth values.'' [{\it Op. cit.}, p. 2] They then proceed to consider two property ascriptions, the MPAC: ``If a system is in an eigenstate of an operator with eigenvalue $v$, then the system has the qualitative property corresponding to such value of the observable.'' And the PPA: ``When in a superposition, the system does have the properties related to the vectors forming the superposition, and they are contradictory.'' [{\it Op. cit.}, p. 2] Arenhart and Krause then argue that: ``when the conditions for application of the minimal principle [MPAC] are met, both states have opposite truth values. But the job is still not done: we must still grant that one of those propositions must always be the case (being so that the other one will be false), as the semantic requirement [SR] for a contradiction seems to demand.'' [{\it Op. cit.}, p. 5] Their conclusion is then the following: 

\begin{quotation}
\noindent {\small``[...] it seems that the semantic requirement [SR] that one of the two terms in a superposition must always be the case (so that we can have a contradiction) is in fact in conflict with the paraconsistent property attribution [PPA] principle. For the latter principle to apply, in the case of a superposition, both `up' and `down' would {\it have to be the case simultaneously.} Recall what happens in the case of the two slit or Schr\"odinger's cat: according to this proposal [PAQS], the particle must go by both slits, the cat must be dead and alive. So, there cannot be alternate truth values in this case, for both must be simply true. So, there is a conflict of the paraconsistent property attribution [PPA] principle with the very requirement that the vectors in a superposition stand for contradictory properties, at least according to the usual semantic requirements [SR] related to contradictions, as it appears in the traditional analysis of this concept. It seems that one cannot have both the claim that $u_x$ [$| \uparrow_x > < \uparrow_x |$] and $d_x$ [$| \downarrow_x > < \downarrow_x |$] are contradictory and the claim that a superposition involves contradictions, as supplied by the paraconsistent property attribution [PPA] principle. As it stands, it seems, these demands are incompatible.'' [{\it Op. cit.}, p. 5]}
\end{quotation}

Let us analyze the presuppositions for the argument to stand. Both conditions, SR and MPAC, imply an analysis either in terms of actuality or actualization. But as we mentioned above, the notion of contradiction that we mean to put forward must consider the realm of potentiality independently of actuality. If the superposition (as a physical notion) exists in potential realm, such conditions cannot be taken into account for they implicitly assume that what is found out in actuality must be directly referred to superpositions irrespectively of their mode of existence. However, if different modes of existence ---to which such conditions make reference--- are considered, there is plenty of room to take them into account within the PAQS. 

The existence of {\it powers} or {\it capacities} in Nature is a well known subject of debate in metaphysics and philosophy of physics \cite{Cartwright89, Marmodoro10}. An analogy with the measurement of a power can show us why the argument of Krause and Arenhart does not follow irrespectively of the metaphysical considerations of the subject under study. Let us, for the sake of the argument, admit that powers (which are not entities) exist in Nature, in a potential realm. For a more detailed analysis of such interpretations using the notion of power we refer to \cite{deRonde15a, daCostadeRonde15}. There exist contradictory powers such as `putting my hand up' or `putting my hand down' (following the PPA). If I `put my hand up' (in actuality) then everyone who is looking will learn that I posses such a power (as demanded by the MPAC), and at the same time, everyone will have observed that (in actuality) I did not `put my hand down' (only one of the two possibilities will be `true' in the actual realm as required by the SR) ---viceversa, when I `put my hand down'.\footnote{This is an analogous situation to the one expressed in a situation in which we have the superposition $c_{1x} | \uparrow_x \rangle + \ c_{2x} | \downarrow_x \rangle$.} The expression of a power in actuality exposes its existence in exactly the same way we can only see an object when light shines upon it. 

Taking into account the {\it Closed Representational Stance} (Section 1.2), every physical theory determines its own specific conditions to expose the existents of which it talks about. Does the measurement of a power involve a collapse? The answer is no: the expression of the power does not mean that the power has been destroyed nor that other powers have ceased to exist. The fact that I `put my hand up' in actuality does not imply in any way that I will cease to have this power in the future. Do we need actuality to claim that a power exists? The answer is no: I could choose not to raise my hand but nonetheless still claim that the power exists ---in the same way that when I close my eyes I can still claim that the table in front of me exists. As remarked in our {\it Constructive Stance} (Section 1.2), actuality is not necessarily considered as a limit to the representation of reality. This simple example attempts to show that if one moves away from the metaphysics of `entities' and `properties' in the actual realm, there are different ways to think about existence and actualization. The classical physical representation in terms of entities and properties in the actual mode of existence might not be the end of the road.

What is at stake in QM is the meaning itself of existence ---QM does not seem to make reference to an ASA---, this is why we are not committed necessarily to a definite metaphysical stance such as the one implied by the OLR. SR and MPAC are not {\it necessary conditions} for every interpretation of QM that we can think of. They already imply a metaphysical stance in which reality is conceived only in terms of entities in the actual mode of existence. But, as we have shown above, if we include the mode of existence within the conditions themselves, we can certainly take into account SR, MPAC and PPA within the PAQS. By claiming that PPA refers to the potential realm while MPCA refers to actualization (something we have discussed in detail in \cite{RFD14}) and SR to the actual realm, all conditions are met by the PAQS and the problems raised by Arenhart and Krause disappear.\\

\noindent {\bf II. The PAQS obscures the meaning of probability in QM.}

\noindent In several passages (e.g., pp. 4 and 5) Arenhart and Krause seem to claim that PAQS does not allow for a good interpretation of probability in the context of QM. More specifically as they argue on p. 13: 

\begin{quotation}
\noindent {\small``Besides that lack of additional explanatory power or enlightenment on the theory [by the PAQS], there are some additional difficulties here. There is a complete lack of symmetry with the standard case of property attribution in quantum mechanics. As it is usually understood, by adopting the minimal property attribution principle, it is not contentious that when a system is in one eigenstate of an observable, then we may reasonably infer that the system has the property represented by the associated observable, so that the probability of obtaining the eigenvalue associated is 1. In the case of superpositions, if they represented properties of their own, there is a complete disanalogy with that situation: probabilities play a different role, a system has a contradictory property attributed by a superposition irrespective of probability attribution and the role of probabilities in determining measurement outcomes. In a superposition, according to the proposal we are analyzing, probabilities play no role, the system simply has a given contradictory property by the simple fact of being in a (certain) superposition.''}
\end{quotation}

Some remarks go in order. Firstly, QM does not only talk about probability. The KS theorem, which we have mentioned before does not talk about probabilities but about actual definite values of observables. As a mater of fact, if QM would be only talking about probability, then there would be no single interpretational problem for the average values of all observables (commuting or not) are perfectly well defined in the theory. The problem is that QM does not describe a mere ensemble of individuals. We do not know what is an individual according to QM (see for discussion \cite{daCostadeRonde14, FrenchKrause}), neither is it clear what is the relation between preexistence and observation. As we know, it makes no sense to assume that the measurement exposes an already preexistent ASA as a known paper of Peres express it, in QM ``unperformed experiments have no results.'' \cite{Peres}. (This is why MPAC needs to be so weak in the first place!) We do not know what it means that a superposition exists and we do not know how to relate it to actual observations. But claiming that the PAQS does not allow for a good interpretation of probability obviously implies that we know what quantum probability is talking about. This is simply not true. We do not know what quantum probability means in terms of a physical concept. We do have a physical interpretation for classical (Kolmogorovian) probability, but this is not the case in QM. There is a whole literature regarding this point \cite{Redei12}, but we can neither forget that the well known interpretations put forward by Popper, with his propensity interpretation of probabilities \cite{Popper82}, and Bohm, with his causal interpretation, were specifically designed in order to find an answer to the meaning of quantum probability.\footnote{As remarked by Bohm \cite[p. 465]{Bohm53} himself: ``[...] in the usual interpretation two completely different kinds of statistics are needed. First, there is the ordinary statistical mechanics, which treats of the distortion of systems among the quantum states, resulting from various chaotic factors such as collisions. The need of this type of statistics could in principle be avoided by means of more accurate measurements [...]. Secondly, however, there is the fundamental and irreducible probability distribution, $P(x)=|\psi(x)|^{2}$ [...]. The need of this type of statistics cannot even in principle be avoided by means of better measurements, nor can it be explained in terms of the effects of random collision processes."} The phrase ``QM is a probabilistic theory", commonly used within the literature (and also addressed by Arenhart and Krause), is from our perspective: either an obvious mathematical statement with no interest ---it only states the well known fact that in QM there is a (non-Kolmogrovian) probability measure assigned via Gleason's theorem--- or a meaningless physical statement, since we do not know what quantum probability is in terms of a physical concept. This was a fact already known to the founding fathers of the theory. For example, as noticed by Schr\"odinger \cite[p. 115]{Bub97} in a letter to Einstein:

\begin{quotation}
\noindent {\small ``It seems to me that the concept of [physical] probability is terribly mishandled these days. [Physical] probability surely has as its substance a statement as to whether something {\small {\it is}} or {\small {\it is not}} the case ---an uncertain statement, to be sure. But nevertheless it has meaning only if one is indeed convinced that the something in question quite definitely {\small {\it is}} or {\small {\it is not}} the case. A [physical] probabilistic assertion presupposes the full reality of its subject.''}
\end{quotation}

\noindent The problem with probability in QM is that due to the formalism there are serious inconveniences to assert the full reality of the subject, in terms of an ASA. Finally, we should remark that Arenhart and Krause mix statements regarding the probability of obtaining one of the terms in a superposition, before and after the measurement has taken place (see e.g. pp. 4 and 10). As we shall argue in {\bf IV}, one needs to be very careful regarding such analysis for it is also well known that conditional probability does not entail a necessary interpretation in terms of a collapse.\footnote{As a matter of fact, the conditional probability of obtaining $B$ at $t_2$ given that $A$ was observed at $t_1$, $p(B, t_2|A, t_1)$, does not relate in any way the probabilities of measuring $A$ at $t_1$, $p(A, t_1)$, and the probability of measuring $B$ at $t_2$, $p(B, t_2)$. I am indebted to Prof. Dieks for pointing out to me this subtle point of non-collapase interpretations. See for discussion: \cite{Dickson98}.}\\

\noindent{\bf III. The PAQS does not explain the MP.}

\noindent Analyzing the PAQS, Arenhart and Krause [{\it Op. cit.}, p. 6] also claim the following:

\begin{quotation}
\noindent {\small ``PAQS makes it even more difficult to understand how a typical measurement of a system in superposition yields always determinate results, but not contradictory results: one must be able to explain how a property possessed by the system disappears, while the other one remains."}
\end{quotation}

\noindent This remark is focused on the MP and the explanation of the phenomenon in the actual realm of existence. As we have argued above (Section 2), once we consider two realms, the problem of actualization cannot be posed exclusively in terms of actuality. Furthermore, Arenhart and Krause claim that the terms which do not get actualized after measurement suddenly ``disappear'', an interpretational move with respect to the PP which is not self evident and presupposes a (physical) collapse. We will come back to this interpretational maneuver in {\bf VI}.\\

\noindent{\bf IV. Contrary to what is claimed by PAQS, contradictions are not observed in QM.}

\noindent The meaning of observation is of course a subtle point in QM, one at the center of all interpretational problems. An important consequence of accepting the CMLR is that ---through the {\it Closed Representational Stance}--- every new theory determines a new experience. Following this line of thought we have criticized the idea that observation in QM must be necessarily considered in terms of classical space-time experience. In this respect, Arenhart and Krause have claimed ---focusing once again on the MP--- that: ``For another disanalogy [of PAQS] with the usual case, one does not expect to observe a system in such a contradictory state: every measurement gives us a system in a particular state, never in a superposition.'' \cite[p. 10]{ArenhartKrause14}

According to our stance, in order to have a closed experience such as that provided, for example, by classical physics, there is a need of coherency between the mathematical structure, the concepts of the theory and the experience exposed through them. Thus when we say that according to Newtonian mechanics a cup falls to the floor accelerated at $9.8 \frac{m}{s^2}$ in $t_1$ seconds, there is a coherency of the statement, of the concepts implicitly used (object, space, time, etc.), the formal prediction (according to the equation of motion) and experience itself. In QM, because we do not have a coherent language that makes contact with the formal structure, experience is not really well defined. Instead, what we have is a weird discourse which constantly contradicts itself. When we talk about ``quantum particles'', we know that they are not particles. Some people argue that ``this is just a way of talking''. As we have attempted to show in this paper, language and, more specifically, physical concepts determine a definite perspective regarding problems and their solutions articulating our possibilities to think about experience (see for discussion \cite{deRondeMassri14}.

Another important aspect regards the fact that one cannot simply ``observe contradictions'' (see for discussion: \cite[pp. 104-105]{deRonde14}). `Contradictions', like `identity' or `causality', are not something that we {\it find outside} in the world; they are instead the basic  metaphysical presuppositions that shape our theories in order to comprise and make sense of experience. The fact that contradictions are observed, or not, needs to be addressed from the standpoint of a coherent interpretation of QM, something we still do not have. As a matter of fact, we still do not know what expresses, according to QM, a `click in a detector'.\\

\noindent{\bf V. The PAQS inflates unnecessarily the population of the world with contradictions.}

\noindent According to Arenhart and Krause [{\it Op. cit.}, p. 10]:

\begin{quotation}
\noindent {\small ``[...] when one takes into account other virtues of a metaphysical theory, such as economy and simplicity, the paraconsistent approach seems to inflate too much the population of our world. In the presence of more economical candidates doing the same job and absence of other grounds on which to choose the competing proposals, the more economical approaches take advantage. Furthermore, considering economy and the existence of theories not postulating contradictions in quantum mechanics, it seems reasonable to employ PriestÕs razor ---the principle according to which one should not assume contradictions beyond necessity (see Priest [12])--- and stick with the consistent approaches. Once again, a useful methodological principle seems to deem the interpretation of superposition as contradiction as unnecessary.''}\end{quotation}

On the one hand, Arenhart and Krause claim that there are more economical interpretations which do exactly the same job the PAQS does. But what is the job done by the PAQS? The PAQS brings forward the possibility to interpret all terms in the superposition as physically existent ---supporting the fact that all terms can be `described', `put to interact' and `be predicted' through the Born rule and the Schr\"odingier equation. The price the PAQS might need to pay is giving up the equation: Actuality = Reality. Thus, the PAQS allow us to investigate the possibility of considering a contradictory (potential) realm independent of actuality. Are there many interpretations that do this job? It is not clear that such is the case. 

As a matter of fact, most interpretations of QM do not even consider quantum superpositions as physical existents (see for a detailed analysis \cite{deRonde15b}). For example, the so called Copenhagen interpretation remains agnostic with respect to the mode of existence of properties {\it prior} to measurement. The same interpretation is endorsed by van Fraassen in his Copenhagen modal interpretation.\footnote{According to van Fraassen \cite[p. 280]{VF91}: ``The interpretational question facing us is exactly: in general, which value attributions are true? The response to this question can be very conservative or very liberal. Both court later puzzles. I take it that the Copenhagen interpretation ---really, a roughly correlated set of attitudes expressed by members of the Copenhagen school, and not a precise interpretation--- introduced great conservatism in this respect. Copenhagen scientists appeared to doubt or deny that observables even have values, unless their state forces to say so. I shall accordingly refer to the following very cautious answer as the {\it Copenhagen variant} of the modal interpretation. It is the variant I prefer.''} In Dieks' realistic modal version, only one of terms is real (actual), while all other terms are considered as possible (in the classical sense). Bohmian versions deny right from the start the existence of quantum superpositions and claim instead the existence of a quantum field that governs the evolution of particles. One might argue that some interpretations, although not explicitly, leave space to consider superpositions as existent in a potential, propensity, dispositional or latent realm. The Jauch and Piron School, Popper or Margenau's interpretations, are a clear example of such proposal (see for discussion \cite{deRonde11} and references therein). However,  within such interpretations the collapse is accepted and potentialities, propensities or dispositions are only defined in terms of `their becoming actual' ---mainly because, forced by the OLR, they have been only focused on providing an answer to the MP. In any case, such realms are not further articulated. Only the many worlds interpretation goes as far as claiming that all terms in the superposition are real in actuality. However, the quite expensive metaphysical price to pay is to argue that there is a multiplicity of unobservable Worlds (branches) in which each one of the terms is actual. 

The PAQS does the job of allowing a further formal development of a realm in which superpositions exist, regardless of actuality.\footnote{Although we believe there is plenty of room to use the PAQS in many interpretations of QM, the author of this paper has argued elsewhere in favor of a non-collapse interpretation which considers the potential realm completely independent of actuality. We will discuss the particular relation of this interpretation to PAQS in \cite{daCostadeRonde14}.} In the sense just discussed the PAQS opens possibilities of development which have not yet been fully investigated. It should be also clear that we are not claiming that all terms in the superposition are actual ---as in the many worlds interpretations--- overpopulating existence with unobservable actualities. What we claim is that PAQS opens the door to consider all terms as existent in potentiality ---independently of actuality. We claim that just like we need all properties to characterize a physical object, all terms in the superposition are needed for a proper characterization of what exists according to QM. We do not believe that this is overpopulating metaphysically the world with contradictions, but rather an attempt to take into account what both the formalism of QM as well as physical experience in the laboratory seems to be telling us. Finally, it is important to remark that ---as we discussed in {\bf IV}--- such contradictory potentialities are observable just in the same way as actual properties can be observed in an object. Potentialities can be observed through actual effectuations in analogous fashion to physical objects ---we never observe all perspectives of an object simultaneously, instead, we observe at most a single subset of actual properties.\\

\noindent{\bf VI. The PAQS does not explain the vanishing of terms in the superposition after measurement.}

\noindent Finally, according to Arenhart and Krause [{\it Op. cit.}, pp. 10-11]: 

\begin{quotation}
\noindent {\small ``[...] a new problem is created by this interpretation [PAQS], because besides explaining what is it that makes a measurement give a specific result when the system measured is in a superposition (a problem usually addressed by the collapse postulate, which seems to be out of fashion now), one must also explain why and how the contradictory properties that do not get actualized vanish. That is, besides explaining how one particular property gets actual, one must explain how the properties posed by the system that did not get actual vanish.''}\end{quotation}

\noindent Arenhart and Krause seem to assume that once the populated world of contradictions is measured all terms except one suddenly disappear. This implies obviously the interpretation of the PP in terms of a collapse (i.e., a physical interaction) of the quantum wave function. But, it is well known that such collapse interpretation is not necessarily the only possible interpretation of the PP. We are inclined to assume a non-collapse interpretation of PP while still considering the specificity of the actualization process in QM (see for discussion \cite{RFD14}). Thus PAQS might allow us to claim that the superpositions remain existent (in potentiality) independently of their measurement (in actuality). We should remark that this is also an important point for modal interpretations. As remarked by P. Vermaas  \cite[p. 295]{Vermaas99}: ``In modal interpretations the state is [...] not updated if a certain state of affairs becomes actual. The non-actualized possibilities are not removed from the description of a system and this state therefor codifies not only what is presently actual but also what was presently possible. These non-actualized possibilities can, as a consequence, in principle still affect the course of later events.''

\section{Final Remarks}

Arenhart and Krause have also called the attention to the understanding of contradiction via the Square of Opposition. Elsewhere, together with Domenech and Freytes, we have also analyzed via the Square of Opposition the meaning of quantum possibility. We argued that the notion of possibility would need to be discussed in terms of the formal structure of the theory itself and that, in such case, one should not study the Classical Square of Opposition but rather an Orthomodular Square of Opposition such as the one explicitly developed in \cite{FRD12}. In  \cite{RFD14b} we provided an interpretation of the Orthomodular Square of Opposition in terms of the notion of potentiality. In a future paper \cite{FRD14} we plan to analyze the proposal of Arenhart and Krause and discuss the meaning of contradiction relating our Orthomodular Square of Opposition with the constraints implied in the MKS theorem. We expect that this analysis will provide us with a better understanding of contradictions in QM. 

As we attempted to show the criticisms of Arenhart and Krause, either arise from assuming the OLR ---to which we are not committed--- or by presupposing certain aspects ---e.g., a collapse interpretation of the PP, an actualist understanding of reality, etc.--- that we have never assumed in the first place. Although the author of this paper has a definite position with respect to the interpretation of QM, we prefer to leave open the PAQS to be used by any interpretation of QM.

PAQS allow us to consider a contradictory realm in which all terms of the superpositions preexist to measurement. In turn this may allow us to provide a physical interpretation in accordance to the latest technical and experimental developments that are taking place today (e.g., quantum computation, quantum teleportation, quantum information processing, etc.) which use the interaction of all terms in the superposition irrespectively of the actual measurement outcome.\footnote{In this respect, the PAQS goes in line, for example, with the ongoing research of the quantum group at Oxford University Computing Laboratory directed by Abramsky and Coecke and the projects about quantum interaction directed by Smets at Amsterdam University.} It is important to remark that some new paraconsistent formalizations of quantum superpositions ---which go in line with the proposal presented in \cite{daCostadeRonde13}--- are being developed \cite{daCostadeRonde14, KrauseArenhart14}.

We believe that science is about confronting the unknown, it implies a humble attitude with respect to experience and a critical understanding of the presuppositions we are willing to make. Science implies the creation and production of new ways of understanding reality; it is not about trying to justify that which we already know. We need to become again a child and observe with admiration and surprise the world that surrounds us, we need to imagine beyond the limits of the impossible, we need to think the unthinkable. While we want to be very cautions about what we know and be (sometimes) very wild about what we do not know, the OLR seems to desire exactly the opposite. But exactly because science has always taken advantage from opposite views and perspectives, we believe that our approach and line of research, although still speculative and in early stages of development, deserves the chance of being further developed.

\section*{Acknowledgements} 

We want to thank Nahuel Sznajderhaus for a careful reading of an earlier draft of this paper. This work was partially supported by the following grants: FWO project G.0405.08 and FWO-research community W0.030.06. CONICET RES. 4541-12 (2013-2014).

\end{document}